\documentstyle[ltwol]{article}

\def\bea{\begin{eqnarray}}
\def\eea{\end{eqnarray}}

\begin{document}
\def\lsim{\mathrel{\lower2.5pt\vbox{\lineskip=0pt\baselineskip=0pt
\hbox{$<$}\hbox{$\sim$}}}}
\def\gsim{\mathrel{\lower2.5pt\vbox{\lineskip=0pt\baselineskip=0pt
\hbox{$>$}\hbox{$\sim$}}}}

\def\gs{SU(2)_{\rm L} \times U(1)_{\rm Y}}
\def\mt{{\tilde{m}^2}_{{\scriptscriptstyle t}}}
\def\mb{{\tilde{m}^2}_{{\scriptscriptstyle b}}}
\def\mz{{m}_{\scriptscriptstyle Z}^{2}}
\def\at{{A_{\scriptscriptstyle t}}}
\def\ab{{A_{\scriptscriptstyle b}}}
\def\qt{{Q_{\scriptscriptstyle t}}}
\def\qb{{Q_{\scriptscriptstyle b}}}
\def\tu{{\tilde{t}_{{\scriptscriptstyle 1}}}}
\def\td{{\tilde{t}_{{\scriptscriptstyle 2}}}}
\def\bu{{\tilde{b}_{{\scriptscriptstyle 1}}}}
\def\bd{{\tilde{b}_{{\scriptscriptstyle 2}}}}
\def\tauu{{\tilde{\tau}_{{\scriptscriptstyle 1}}}}
\def\taud{{\tilde{\tau}_{{\scriptscriptstyle 2}}}}
\def\mnu{{\tilde{m}^2_{{\scriptscriptstyle \nu}}}}
\def\mtau{{\tilde{m}^2}_{{\scriptscriptstyle \tau}_{\scriptscriptstyle 1}}}
\def\mtad{{\tilde{m}^2}_{{\scriptscriptstyle \tau}_{\scriptscriptstyle 2}}}
\def\w{{\tilde{W}^{\pm}}}
\def\h{{\tilde{H}^{\pm}}}
\def\sw{s_{{\scriptscriptstyle W}}}
\def\cw{c_{{\scriptscriptstyle W}}}
\def\mfa{{\tilde{m}^2}_{{\scriptscriptstyle fa}}}
\def\mfb{{\tilde{m}^2}_{{\scriptscriptstyle fb}}}
\def\mfc{{\tilde{m}^2}_{{\scriptscriptstyle fc}}}
\def\mfd{{\tilde{m}^2}_{{\scriptscriptstyle fd}}}
\def\mi{{\tilde{M}^2}_{{\scriptscriptstyle i}}}
\def\mj{{\tilde{M}^2}_{{\scriptscriptstyle j}}}
\newcommand{\sfe}{{\tilde f}}
\newcommand{\sne}{{\tilde \chi^o}}
\newcommand{\scp}{{\tilde \chi^+}}
\newcommand{\dx}{dx}  
\newcommand{\dy}{dy}  
\newcommand{\slas}[1]{\rlap/ #1}
\newcommand{\diag}{{\rm diag}}
\newcommand{\Tr}{{\rm Tr}}
\title{Do heavy sfermions decouple from low energy Standard Model?}

\author{ANTONIO DOBADO}

\address{Departamento de F{\'\i}sica Te{\'o}rica, Universidad Complutense de Madrid,
 28040-- Madrid, Spain\\E-mail: dobado@eucmax.sim.ucm.es}   

\author{MARIA J. HERRERO  and  SIANNAH PE{\~N}ARANDA}

\address{Departamento de F{\'\i}sica Te{\'o}rica, Universidad Aut{\'o}noma de Madrid,
  Cantoblanco, 28049-- Madrid, Spain\\E-mail: herrero@delta.ft.uam.es,\\ 
  siannah@delta.ft.uam.es}  

\twocolumn[\vspace*{-1cm}
\begin{flushright}
{June 1998}\\
{FTUAM 98/11}\\
{hep-ph/9806488}
\end{flushright}
\maketitle\abstracts{ We explore analytically how does the Standard Model emerge 
as the quantum low energy effective theory of the Minimal Supersymmetric 
Standard Model (MSSM) in the decoupling limit where the sparticles are much 
heavier than the electroweak scale. In this work we integrate the 
sfermions to one-loop and compute their contributions to the effective 
action for standard electroweak gauge bosons. A proof of 
decoupling of sfermions is performed by analyzing the resulting effective 
action in the asymptotic limit $m_{\tilde f} \gg m_z$. A discussion on 
how the decoupling takes place in terms of both the sparticle physical 
masses and the non-physical MSSM parameters is included.}]

\section{Introduction}
\label{sec:mssm}
\renewcommand\baselinestretch{1.3}
\hspace*{0.5cm} The Standard Model (SM) is a pillar of success as an effective theory. Experimental 
measurements agree with SM radiative corrections to a precision of greater than $0.1\%$. However, SM
contains nagging theoretical problems which cannot be solved without the introduction of some new 
physics. In this sense, supersymmetry (SUSY) is the favorite of many theorists.
The simplest model 
of this type is called the Minimal Supersymme\-tric Standard Model (MSSM)~\cite{HAMSSM}, which is the 
one we have chosen to work with in this paper.

One interesting aspect that arises in these softly broken SUSY theories, and 
in particular in the MSSM, is the question of decoupling of heavy sparticles 
from the low energy  SM and how does it really occurs if it occurs at all. We will concentrate our 
attention in this subject. In general, perturbative considerations~\cite{OJO} lead us to believe that heavy
particles can be decoupled from low energy degrees of freedom. We expect that the lagrangian 
des\-cribing the low energy degrees of freedom is affected by heavy particles only through renormalization
effects and higher dimension operators which become negligible as the particles are made infinitely massive.

At present, there are indications that when the spectrum of supersymmetric particles 
at the MSSM is considered
much  heavier  than the low energy electroweak scale they decouple from the low energy physics, even at the quantum level, and 
the resulting low energy effective theory is the SM itself. However, a rigorous proof 
of decoupling is still lacking. On one hand there are numerical studies of 
observables that measure electroweak radiative corrections, like $\Delta r$ and $\Delta \rho$
 \cite{CHA}, or the $S, T$ and $U$ parame\-ters~\cite{HA2} as well as in the $Z$ boson, top quark and Higgs 
decays \cite{SOLA}, which indicate  that the one loop corrections from supersymmetric particles 
decrease up to negligible values in the limit of very heavy sparticle masses. Decoupling of SUSY
particles is also found in some analytical studies of these and related 
observables~{\cite{CHA}-\cite{GL1}}.

It has been known for some time that there are some exceptions where the 
Decoupling Theorem~\cite{OJO} does not apply. Particularly interesting are the cases of the Higgs particle and 
the top quark in the SM which are known not to decouple from low energy physics \cite{VE,CHI,MJ}.
The question whether the Decoupling Theorem applies or not in the case of heavy 
sparticles in MSSM is not obvious at all, in our opinion. The MSSM is a gauge theory which
incorporates the spontaneous symmetry breaking $\gs \rightarrow U(1)_{\rm em}$ and chiral 
fermions as the SM and  therefore, the direct application of this theorem should, in the
principle, be questioned~\cite{ANT}.

In our opinion, a formal proof of decoupling must involve the explicit 
computation of the effective action by integrating out one by one all the 
sparticles in the MSSM to all orders in perturbation theory, and by considering 
the heavy sparticle masses limit. The proof will be conclusive if the remaining   
effective action, to be valid at energies much lower than the supersymmetric particle 
masses, turns out to be that of the SM with all the SUSY effects being absorbed into  
a redefinition of the SM parameters or else they are suppressed by inverse powers  
of the SUSY particle masses and vanish in the infinite masses limit.

In this work we discuss part of the effective action which results by integrating out the
sfermions of the MSSM at the one loop level. This is a reduced version of the more complete papers
to which we refer the reader for a more detailed discussion~\cite{GEISHA,DMS}. Here, we have devoted 
our attention on the derivation of the two, three and four-point functions with external
$W^{\pm}$, Z and $\gamma$ gauge bosons. In order to keep our computation of the heavy 
SUSY particle quantum effects in a general form we have chosen to work with the masses 
themselves. Nevertheless, a discussion on how the decoupling takes places in terms of both the physical 
sparticle masses and the non-physical MSSM parameters, as the $\mu$-parameter or the 
soft-SUSY-breaking parameters, 
$M_{{\scriptscriptstyle {\tilde Q}}}, M_{{\scriptscriptstyle {\tilde U}}}, 
M_{{\scriptscriptstyle {\tilde D}}}, M_{{\scriptscriptstyle {\tilde L}}},
M_{{\scriptscriptstyle {\tilde E}}}$, is included.

It is important to remark that we have considered the physically plausible situation where all 
the sparticle masses are large as compared to 
the electroweak scale but they are allowed, in principle, to be different from each other. 
We will explore the interesting question of to what extent the usual hypothesis of SUSY masses being  
gene\-rated by soft-SUSY-breaking terms and the universality of the mass parameters
do or do not play a relevant role  in getting decoupling. In fact, we will show in this
paper, that the basic requirement of $\gs$ gauge invariance on the SUSY breaking 
terms is sufficient to obtain decoupling in the MSSM.

Finally, we would like to point out that in order to evaluate
analytically the large SUSY masses limit of the Green functions
we have applied the so-called m-Theorem \cite{GMR}, which provides a rigorous technique to compute 
Feynman integrals with both large and small masses in the asymptotic regime of the large 
masses being very heavy. This theorem will enable us not only to disregard integrals that do not
contribute to the limit of large masses without having to compute them, but also 
to eva\-luate exactly the non-decoupling contributions.

The paper is organized as follows: In section 2 we present a brief discussion about the 
large mass limit for all the sfermions at the MSSM. The third section is devoted to present 
the effective action for the electroweak gauge bosons  $W^{\pm}$, Z and $\gamma$ in the MSSM 
that results by integrating out, in the path integral, sfermions to one-loop.
The asymptotic results in the large SUSY masses limit for the two, three and four-points functions
are included and analyzed in section 3. Finally, the conclusions are summarized in section 4.

\section{The large supersymmetric masses limit.}
\label{sec:mlimit}
\hspace*{0.5cm} As we point out before, in the present work we are interested in the Green 
functions with external electroweak gauge
bosons and in the large mass limit of the SUSY particles, which means the situation
where all the sparticle masses are much larger than the electroweak scale and the external
momenta. In particular this could be the case if the sparticle masses are well above 
$m_{\scriptscriptstyle Z}, m_{\scriptscriptstyle W}$ and $m_{\scriptscriptstyle t}$ 
but still below the few TeV upper bound that is imposed by the standard solution of the 
hierarchy problem. Furthermore, unless we are in a particular model, the masses of the 
various sparticles are, in general, different and independent. Therefore, we must take these 
masses to be large as compared to the external gauge boson masses and external momenta, but we must 
specify, in addition, how they compare to each other. More specifically, we assume here the most 
plausible situation where all the 
sparticle masses are large but close to each other; namely 
$\tilde{m}_{\scriptscriptstyle i}^{2}, \tilde{m}_{\scriptscriptstyle j}^{2} \gg 
M _{\scriptscriptstyle EW}^{2}, k^{2}$ and
$|\tilde{m}_{\scriptscriptstyle i}^{2}-\tilde{m}_{\scriptscriptstyle j}^{2}| \ll
|\tilde{m}_{\scriptscriptstyle i}^{2}+\tilde{m}_{\scriptscriptstyle j}^{2}|$, where 
$M _{\scriptscriptstyle EW}$ denotes any of the electroweak masses involved 
$(m_{\scriptscriptstyle Z}, m_{\scriptscriptstyle W}, m_{\scriptscriptstyle t},
\ldots)$ and $k$ denotes any external momentum. Notice that this includes the case that has
been the most studied in the literature where universality of sparticle masses is assumed.

In principle, our asymptotic limit is on the physical masses, which implies,
of course, some conditions over the parameters of the model. In other words, 
our masses hypothesis, together with the requirement that all the sparticles must be heavier 
than their corresponding partners, imply some constraints on the SUSY parameters. In
particular, in the squarks sector, if we ignore mixing between different generations to avoid 
unacceptable large flavor changing neutral currents and if we use the notation of the third family
for the mass eigenstates $\tu,\td,\bu,\bd$ and the corresponding mass squared
eigenvalues by $\tilde{m}^2_{{\scriptscriptstyle t}_{1,2}}$,
$\tilde{m}^2_{{\scriptscriptstyle b}_{1,2}}$, it can be shown the following
constraints on the soft SUSY breaking and $\mu$ parameters hold\cite{GEISHA}:
\bea 
\label{eq:rest1}
M_{{\scriptscriptstyle {\tilde Q}}}^{2}, M_{{\scriptscriptstyle {\tilde U}}}^{2} 
\gg m_{t}^{2}, m_{{\scriptscriptstyle Z}}^{2}\,\,&,& \hspace*{0.1cm}
|M_{{\scriptscriptstyle {\tilde Q}}}^{2}-M_{{\scriptscriptstyle {\tilde U}}}^{2}| \ll
|M_{{\scriptscriptstyle {\tilde Q}}}^{2}+M_{{\scriptscriptstyle {\tilde U}}}^{2}|\nonumber \\
\noalign{\vskip 3pt}
m_{\scriptscriptstyle t}^{2} (\at-\mu \cot{\beta})^{2} &<& 
M_{{\scriptscriptstyle {\tilde Q}}}^{2} M_{{\scriptscriptstyle {\tilde U}}}^{2}\,.
\eea
\hspace*{0.5cm}Here $\at$ is the trilinear coupling and $\cot{\beta} \equiv v_{1}/v_{2}$. The first 
condition implies, in turn, the limiting behaviour 
$\tilde{m}_{t_{1}}^{2}\rightarrow M_{{\scriptscriptstyle {\tilde Q}}}^{2}$\hspace*{0.1cm},
$\tilde{m}_{t_{2}}^{2}\rightarrow M_{{\scriptscriptstyle {\tilde U}}}^{2}$ \hspace*{0.1cm}. The
second condition means that $M_{{\scriptscriptstyle {\tilde Q}}}$ and 
$M_{{\scriptscriptstyle {\tilde U}}}$ must be close to each other and the third one means 
that the mixing can never be arbitrarily large. Similar conclusions can be obtained for the sbottoms.

In summary, in other to get large stop and sbottom masses one needs large 
values of the SUSY breaking masses $M_{{\scriptscriptstyle {\tilde Q}}},
M_{{\scriptscriptstyle {\tilde U}}}$ and $M_{{\scriptscriptstyle {\tilde D}}}$ as 
compared to the electroweak scale and, in order not to get a too large mixing, the trilinear couplings
 $\at, \ab$ and the $\mu$ parameter must be constrained from above by the previous inequalities.
Notice that an arbitrarily large $\mu$ or $\at, \ab$ with $M_{{\scriptscriptstyle {\tilde Q}}},
M_{{\scriptscriptstyle {\tilde U}}},M_{{\scriptscriptstyle {\tilde D}}}$ fixed is not
allowed. 

We would like to mention that the asymptotic limit considered here is not the unique possibility 
to study decoupling. Other possibilities are now under study~\cite{GEISHA}.

\section{Effective action for the electroweak gauge bosons to one-loop.}
\hspace*{0.5cm} This section is devoted to present the computation of the part of the effective
action that contains the two, three and four-point Green's functions with external gauge bosons,
$A, Z, W^\pm$, which results by integrating out all the sfermions particles of the MSSM at the one
loop level. Details of the computation, including the integration of neutralinos $\sne$ and charginos 
$\scp$ can be found in~\cite{GEISHA,DMS}. The computation has been performed using 
dimensional regularization.

We start by writing, in a general and compact form, the effective action for the standard particles,
$\Gamma_{eff}[\phi]$, that is defined through functional integration of all the
sparticles of the MSSM,
\bea
\label{eq:gammaeff}
{\rm e}^{i\Gamma_{eff}[\phi]}&=&\int [{\rm d}\tilde\phi]\,{\rm e}^{i \Gamma_{\rm MSSM}
  [\phi,\tilde\phi]} \,\,,\nonumber \\
\noalign{\vskip 3pt}
\Gamma_{\rm MSSM}[\phi,\tilde\phi] &\equiv& \int \dx{\cal L}_{\rm
  MSSM}(\phi,\tilde \phi)\,\,;\,\,{\rm d}x\equiv{\rm d}^4x \,,
\eea
where $\phi=l,q,A,W^\pm,Z,g,H$ are the SM particles, $\tilde \phi=\tilde l,\tilde q,\tilde A,
\w,\tilde Z,\tilde g,\tilde H$ their supersymmetric partners, and 
${\cal L}_{\rm MSSM}$ is the lagrangian of the MSSM~\cite{GEISHA}. In the following we will use 
the notation of ref.~\cite{GEISHA}.

As we have already said, in the present work we are interested only in the sfermions contribution.
The corresponding part of the effective action can be written as:
\begin{equation}
\begin{array}{l}
\label{eq:gammaeffF}
\displaystyle 
e^{i \Gamma_{eff}^{\tilde{f}} [{\scriptscriptstyle V}]} = \int [d\tilde{f}] [d\tilde{f}^{*}] 
e^{i \Gamma_{\tilde{f}} [{\scriptscriptstyle V},\tilde{f}]} \,\,,
\end{array}
\end{equation}
where $\tilde{f} = \tilde{q}, \tilde{l} \,\,\,; V=W^\pm, Z, A$ and 
$\Gamma_{\sfe}[V,\sfe]$ is the action for the sfermions.

Notice that the effective action as a function of the n-point Green functions, 
$\Gamma_{\mu \, \nu ... \, \rho}^{V_{1} V_{2} ... V_{n}}$, 
can be written in the following form:
\bea
\label{eq:eff} 
\Gamma_{eff}^{\tilde{f}} [V] &=& \sum_{n} \frac{1}{m!} \int 
 {\rm d}^4x_{1} ... {\rm d}^4x_{n} \,\nonumber \\
&& \hspace*{0.3cm}\Gamma_{\mu \, \nu ... \, \rho}^{V_{1} V_{2} ... V_{n}} (x_{1} \, x_{2} ... \,x_{n}) 
V_{1}^{\mu} \,V_{2}^{\nu} \,... V_{n}^{\rho}\,\,, 
\eea
with $V_{i}\, (i=1... n)$ being the external gauge bosons and $m$ denotes 
the number of these bosons which are identical.

In order to perform the functional integration, it is convenient to write the
classical action in terms of operators. We have computed $\Gamma_{eff}^{\tilde{f}} [V]$ 
by using the standard path integral techniques. The details of the computation 
can be found in \cite{GEISHA}. It is also worth mentioning that we have worked in the momentum
space, which simplify the calculation considerably.

The total resulting effective action for two, three and four-point functions, which are generated 
from sfermions can be summarized in the following expression:
\bea
  \label{eq:totaleff}
\displaystyle 
\Gamma_{eff}^{\tilde{f}} [V] &=& i \Tr ({A_{\tilde{f}}^{(0)}}^{-1} A_{\tilde{f}}^{(2)}) - 
\frac{i}{2} \Tr ({A_{\tilde{f}}^{(0)}}^{-1} A_{\tilde{f}}^{(1)})^{2} \nonumber\\
&-&i \Tr ({A_{\tilde{f}}^{(0)}}^{-1} A_{\tilde{f}}^{(1)} {A_{\tilde{f}}^{(0)}}^{-1} A_{\tilde{f}}^{(2)})
\nonumber\\
&+& \frac{i}{3} \Tr ({A_{\tilde{f}}^{(0)}}^{-1} A_{\tilde{f}}^{(1)})^{3} - 
\frac{i}{2} \Tr ({A_{\tilde{f}}^{(0)}}^{-1} A_{\tilde{f}}^{(2)})^{2}\nonumber\\
&+& i \Tr ({A_{\tilde{f}}^{(0)}}^{-1} A_{\tilde{f}}^{(1)} {A_{\tilde{f}}^{(0)}}^{-1} A_{\tilde{f}}^{(1)}
{A_{\tilde{f}}^{(0)}}^{-1} A_{\tilde{f}}^{(2)})\nonumber\\
&-& \frac{i}{4} \Tr ({A_{\tilde{f}}^{(0)}}^{-1} A_{\tilde{f}}^{(1)})^{4}+O(V^{5})\,\,,
\eea
where the operators are,
\bea
  \label{eq:opersfdef}
  A_{\sfe kp}^{(0)}&\equiv& (2\pi)^4 \delta(p+k) \,(k^2-\tilde M_{f}^2)\,\,, \nonumber\\
  A_{\sfe kp}^{(1)}&\equiv& -(2\pi)^4\int {\rm d\tilde{q}} \, \delta(p+k+q) (q_{\mu}+2p_{\mu})\nonumber\\
  && \left\{eA_{q}^{\mu} \hat{Q}_f +\frac{g}{c_w} Z_{q}^\mu \hat{G}_f+
  \frac{g}{\sqrt{2}} W_{q}^{+\mu} \Sigma_f^{tb}+{\rm h.c.} \right\}\,\,, \nonumber\\
  A_{\sfe kp}^{(2)}&\equiv&(2\pi)^4\int {\rm d\tilde{q}}\, {\rm d\tilde{r}} \, \delta(p+k+q+r) 
  \left\{ e^2 \hat{Q}_f^2 A_{\mu \,q}A_{r}^\mu \right. \nonumber\\
  &+& \frac{2\,g\,e}{c_w}A_{\mu \,q}Z_{r}^\mu
  \hat{Q}_f\hat{G}_f+\frac{g^2}{c_w^2}\hat{G}_f^2 Z_{\mu \,q}Z_{r}^\mu\nonumber\\
  &+& \frac{g^2}{2}\Sigma_f W_{\mu \,q}^{+} W_{r}^{\mu-}+
  \frac{eg}{\sqrt{2}} Y_{\tilde{f}}A_{\mu \,q}\left(W_{\mu \,r}^+ \Sigma_f^{tb}+
  W_{\mu \,r}^- \Sigma_f^{bt}\right)\nonumber\\
  &-& \left. \frac{g^2}{\sqrt{2}}Y_{\tilde{f}}\frac{s_w^2}{c_w}
  Z_{\mu \,q}\left(W_{\mu \,r}^+ \Sigma_f^{tb}+W_{\mu \,r}^- \Sigma_f^{bt}\right)\right\} \,\,.
\eea

In the above expressions and in the following, $\tilde{f}$ is a four-entries column vector 
including the four mass eigenstates per generation, i.e $(\tu, \td,\bu, \bd)$ for squarks and
$(\tilde{\nu}, 0, \tauu, \taud)$ for sleptons. The sum $\sum_{\tilde{f}}$ is over the three 
generations and, in the case of squarks, it 
runs also over the $N_{c}$ color indexes. The coupling matrices $\hat{Q_{f}}, \hat{G_{f}}, 
\Sigma_{f}^{tb}$ and $\Sigma_{f}$ can be found in~\cite{GEISHA}. The parameter $Y_{\tilde{f}}$ and 
the corresponding mass matrices are given by:
\bea
Y_{\tilde{f}}= \frac{1}{3}\hspace*{0.2cm} 
if \hspace*{0.2cm} \tilde{f} = \tilde{q}\hspace*{0.3cm}or\hspace*{0.3cm}
Y_{\tilde{f}}= -1 \hspace*{0.2cm} &if& \hspace*{0.2cm} \tilde{f} = \tilde{l}\,\,,\nonumber\\
\tilde{M}_{f}^{2} = \diag(\tilde{m}_{t_{1}}^{2},\tilde{m}_{t_{2}}^{2},
\tilde{m}_{b_{1}}^{2},\tilde{m}_{b_{2}}^{2})\hspace*{0.35cm} 
&if& \hspace*{0.35cm} \tilde{f} = \tilde{q}\,\,,\nonumber\\
\tilde{M}_{f}^{2} = \diag(\tilde{m}_{\nu}^{2},0,\tilde{m}_{\tau_{1}}^{2},
\tilde{m}_{\tau_{2}}^{2})\hspace*{0.35cm} 
&if& \hspace*{0.35cm} \tilde{f} = \tilde{l}\,\,.
\eea

Clearly, we can identify the first and second terms in eq.~(\ref{eq:totaleff}) with the one-loop
contributions to the two-point functions, the third and fourth terms with the contributions to the
three-point function and the last three terms are the corresponding contributions to the four-point functions.

In order to get the explicit expressions for the two-point functions one must work 
out the traces in the above formulae. Basically one must substitute all the operators 
and compute all the appearing Dirac traces. The traces also 
involve to perform the sum in the corresponding matrix indexes, the sum over the various 
types of sfermions and the sum in color indexes in the case of squarks. We have done this 
computation, in addition, by diagramma\-tical methods
and we have found the same results.

In the following we will present the results of the two, three and four-point functions
with external gauge bosons at one loop.

\subsection{Effective action for the two-point functions.}
\hspace*{0.5cm}We present here the contributions to the two-point function in momentum space, 
$\Gamma^{V_{1}\, V_{2}}_{\mu\, \nu}$. They are given
in terms of the coupling matrices and the one-loop integrals:
\bea
\label{eq:gammaAA} 
\displaystyle \Gamma^{A\, A}_{\mu\, \nu}(k) &=& 
{\Gamma_0}^{A\, A}_{\mu\, \nu}(k)
+i e^{2} \sum_{\tilde{f}} \left\{ 2 \sum_{a} 
(\hat{Q}_{f}^{2})_{aa} I_{0}\, g_{\mu \nu}\right.\nonumber\\
&& - \left.\sum_{ab} (\hat{Q}_{f})_{ab} 
(\hat{Q}_{f})_{ba} I_{f_{\mu \nu}}^{ab}\right\}\\
\label{eq:gammaZZ} 
\displaystyle \Gamma^{Z\, Z}_{\mu\, \nu}(k) &=& 
{\Gamma_0}^{Z\, Z}_{\mu\, \nu}(k)
+i \frac{g^{2}}{c_{{\scriptscriptstyle W}}^{2}} 
\sum_{\tilde{f}} \left\{ 2 \sum_{a}  
(\hat{G}_{f}^{2})_{aa} I_{0}\,g_{\mu \nu} \right.\nonumber\\
&& \left.- \sum_{ab} (\hat{G}_{f})_{ab} 
(\hat{G}_{f})_{ba} I_{f_{\mu \nu}}^{ab}\right\}\\ 
\label{eq:gammaAZ} 
\displaystyle \Gamma^{A\, Z}_{\mu\, \nu}(k) &=&\Gamma^{Z\, A}_{\mu\, \nu}(k)= 
\frac{i g e}{\cw} \sum_{\widetilde{f}} \left\{ 
2 \sum_{a} (\widehat{Q}_{f} \widehat{G}_{f})_{a\, a}I_{0}\, 
g_{\mu\, \nu}\right.\nonumber\\
&&- \left.\sum_{ab}  (\widehat{Q}_{f})_{a\, b} (\widehat{G}_{f})_{b\, a} 
I^{a\,b}_{f_{\mu\, \nu}} \right\}\\ 
\label{eq:gammaWW} 
\displaystyle \Gamma^{W\, W}_{\mu\, \nu}(k) &=&
{\Gamma_0}^{W\, W}_{\mu\, \nu}(k)
+ \frac{i g^2}{2} \sum_{\widetilde{f}}  \left\{ \sum_{a} (\Sigma_{f})_{a\, a}
I_{0}\, g_{\mu\, \nu}\right.\nonumber\\
&& \left.-\sum_{a,b} (\Sigma_{f}^{t\,b})_{a\, b} 
(\Sigma_{f}^{t\,b})_{a\, b} I^{a\,b}_{f_{\mu\, \nu}}\right\} 
\eea
In the above formulae the indexes $a$ and $b$ run from one to four, corresponding to the 
four entries of the
column vector $\tilde{f}$.  ${\Gamma_0}^{V\, V}_{\mu\, \nu}$ $(V = Z, W)$ and 
${\Gamma_0}^{A\, A}_{\mu\, \nu}$ are the two-point functions at tree level, which 
are defined by:
\bea
\label{eq:gam}
{\Gamma_0}^{V\, V}_{\mu\, \nu} (k) &=& (M_{\scriptscriptstyle V}-k^{2}) g_{\mu\, \nu} + 
\left(1 - \frac{1}{\xi_{{\scriptscriptstyle V}}}\right) k_{\mu} k_{\nu} \,\,;\nonumber\\
{\Gamma_0}^{A\, A}_{\mu\, \nu} &=& -k^{2}g_{\mu\, \nu} + 
\left(1 - \frac{1}{\xi_{{\scriptscriptstyle A}}}\right) k_{\mu} k_{\nu} \,\,, 
\eea

The one-loop integrals $I_{0}(\mfa)\,, I^{a\,b}_{f_{\mu\, \nu}}(k, \tilde{m}_{f_{a}}, 
\tilde{m}_{f_{b}})$ are defined in Appendix A.

\subsection{Three and four-point functions.}
\hspace*{0.5cm}For simplicity, we will show here the results for the three and four-point functions in the more general and compact form. The corresponding effective action for the three and four-point functions, 
${\Gamma_{eff}^{\tilde{f}} [V]}_{[3]}$ and ${\Gamma_{eff}^{\tilde{f}} [V]}_{[4]}$ can be expressed as,
\bea
\label{eq:eff3}
&&{\Gamma_{eff}^{\tilde{f}} [V]}_{[3]} =i(2\pi)^4 \int {\rm d\tilde{p}} \,{\rm d\tilde{q}} 
 \,{\rm d\tilde{k}}\,\, \delta(p+k+q) \nonumber\\
&& \hspace*{0.3cm}\sum_{\tilde{f}}\left(\sum_{a,b} (\hat{O}_{\,p}^{1\,\mu})_{ab} (\hat{O}_{\,\,q k}^{2\,\nu\sigma})_{ba}
  T^{a\,b}_{\mu}\,g_{\nu \sigma}\right. \nonumber\\ 
&& \hspace*{0.1cm}\left. -\frac{1}{3} \sum_{a,b,c} (\hat{O}_{p}^{1\,\mu})_{ab}
(\hat{O}_{\,q}^{1\,\nu})_{bc} (\hat{O}_{\,k}^{1\,\sigma})_{ca} \,\,
  T^{a\,b\, c}_{\mu \,\nu \,\sigma}\right), \nonumber\\
\\
 \label{eq:eff4} 
&&{\Gamma_{eff}^{\tilde{f}}[V]}_{[4]} =-i (2\pi)^4 \int {\rm d\tilde{p}} \,{\rm d\tilde{q}} 
 \,{\rm d\tilde{k}} \,{\rm d\tilde{r}} \,\,\delta(p+k+q+r) \nonumber\\
 && \hspace*{0.3cm}\sum_{\tilde{f}}\left( \frac{1}{2}\sum_{a,b} (\hat{O}_{\,\,p q}^{'2\,\mu\nu})_{ab} 
 (\hat{O}_{\,\,q k}^{'2\,\sigma\lambda})_{ba} \,g_{\mu\,\nu}g_{\sigma\lambda}\,
 J^{a\,b}\right. \nonumber\\ 
&& \hspace*{0.2cm} -\sum_{a,b,c}(\hat{O}_{p}^{1\,\mu})_{ab} (\hat{O}_{\,q}^{1\,\nu})_{bc}
(\hat{O}_{\,\,q k}^{'2\,\sigma\lambda})_{ca}\,g_{\sigma \lambda} \,\,
J^{a\,b\, c}_{\mu \,\nu} \nonumber\\
  && \left.+\frac{1}{4} \sum_{a,b,c,d}(\hat{O}_{p}^{1\,\mu})_{ab}
(\hat{O}_{\,q}^{1\,\nu})_{bc} (\hat{O}_{\,k}^{1\,\sigma})_{cd}
(\hat{O}_{\,r}^{1\,\lambda})_{da} \,\, J^{a\,b\,c\,d}_{\mu \,\nu\, \sigma\,\lambda} \right)\,,\nonumber\\
\eea
where, similarly to the two-point functions, the indexes $a, b, c$ and $d$ run from 
one to four and the "operators" $\hat{O}_{\,p}^{1\,\mu}, \hat{O}_{\,\,p q}^{2\,\mu\nu}$ and
$\hat{O}_{\,\,p q}^{'2\,\mu\nu}$ can be summarized by,
\begin{eqnarray}
  \label{eq:operos}
 \hat{O}_{\,p}^{1\,\mu} &=& \left\{eA_{p}^{\mu} \hat{Q}_f +\frac{g}{c_w} Z_{p}^\mu \hat{G}_f+
  \frac{g}{\sqrt{2}} W_{p}^{+\mu} \Sigma_f^{tb}+{\rm h.c.} \right\}\,\,, \nonumber\\
 \hat{O}_{\,\,p q}^{2\,\mu\nu} &=& \left\{ e^2 \hat{Q}_f^2 A_{\mu \,p}A_{q}^\nu+
 \frac{2\,g\,e}{c_w}A_{\mu \,p}Z_{q}^\nu
  \hat{Q}_f\hat{G}_f\right.\nonumber\\
  &+& \left.\frac{g^2}{c_w^2}\hat{G}_f^2 Z_{\mu \,p}Z_{q}^\nu+\frac{g^2}{2}
  \Sigma_f W_{\mu \,p}^{+} W_{q}^{\nu-} \right\} \,\,,\nonumber\\
 \hat{O}_{\,\,p q}^{'2\,\mu\nu} &=& \left\{ \hat{O}_{\,\,p q}^{2\,\mu\nu} +
 \frac{eg}{\sqrt{2}} Y_{\tilde{f}}A_{\mu \,p}\left(W_{\nu \,q}^+ \Sigma_f^{tb}+
  W_{\nu \,q}^- \Sigma_f^{bt}\right)\right.\nonumber\\
  &-& \left.\frac{g^2}{\sqrt{2}}Y_{\tilde{f}}\frac{s_w^2}{c_w}
  Z_{\mu \,p}\left(W_{\nu \,q}^+ \Sigma_f^{tb}+W_{\nu \,q}^- \Sigma_f^{bt}\right)\right\} \,\,.
\end{eqnarray}

$T^{a\,b}_{\mu}, \,T^{a\,b\, c}_{\mu \,\nu \,\sigma}, \,J^{a\,b},
\,J^{a\,b\, c}_{\mu \,\nu}$ and $J^{a\,b\,c\,d}_{\mu \,\nu\, \sigma\,\lambda}$ are the one-loop integrals, which are given in Appendix A.

It is important to emphasize that all these formulae are exact to one loop.

\section{The Green's functions in the large mass limit.}
\label{section:limit}
\hspace*{0.5cm}Since we are interested in the large mass limit of the SUSY particles we need to have at 
hand not just the exact results of the above mentioned integrals but their asymptotic expressions to
be valid in that limit. We have analyzed the integrals by means of the so-called m-Theorem~\cite{GMR}. 
This theorem provides a powerful technique to study the asymptotic behaviour of Feynman 
integrals in the limit where some of the masses are large. Notice that this is not
trivial since some of these integrals are divergent and the interchange of the 
integral with the limit is not allowed. Thus, one should first compute the integrals in 
dimensional regularization and at the end take the large mass limit. Instead of this direct
way it is also possible to proceed as follows: First, in order to decrease the ultraviolet
divergent degree, one rearranges the integrand through
algebraic manipulations up to se\-parate the Feynman integral into a divergent part, which 
can be evaluated exactly using the standard techniques of dimensional regularization, and a convergent part that satisfies
the requirements demanded by the m-Theorem and therefore, goes to zero in the infinite mass
limit. By means of this procedure the correct asymptotic behaviour
of the integrals is guaranteed. Some examples of the computation of the Feynman integrals by means of the m-Theorem as well
as details of this theorem are given in~\cite{GEISHA}. The results for the above one loop integrals
in the large masses limit are presented also in the Appendix A of this paper.

We now proceed to present the asymptotic expressions for the Green's functions
in the large sfermions masses limit. Making use of the results of the one-loop 
integrals given in eqs.(\ref{eq:io}-\ref{eq:last4}) and by using the formulae of 
eqs.(\ref{eq:gammaAA}-\ref{eq:gammaWW}), (\ref{eq:eff3}) and (\ref{eq:eff4}), we 
find the results summarized in the next subsections. 

\hspace*{0.5cm}All the results presented in the following are valid for:
\bea
\hspace*{0.3cm}\tilde{m}_{t_{1}}^{2}, \tilde{m}_{t_{2}}^{2}, 
\tilde{m}_{b_{1}}^{2}, \tilde{m}_{b_{2}}^{2} &\gg& k^{2}\,\,\nonumber\\
|\tilde{m}_{t_{1}}^{2} - \tilde{m}_{t_{2}}^{2}| &\ll&
|\tilde{m}_{t_{1}}^{2} + \tilde{m}_{t_{2}}^{2}| \nonumber\\
|\tilde{m}_{b_{1}}^{2} - \tilde{m}_{b_{2}}^{2}| &\ll& 
|\tilde{m}_{b_{1}}^{2} + \tilde{m}_{b_{2}}^{2}|\,\, {\rm and} \nonumber\\
|\tilde{m}_{t_{i}}^{2} - \tilde{m}_{b_{j}}^{2}| &\ll&
|\tilde{m}_{t_{i}}^{2} + \tilde{m}_{b_{j}}^{2}| \hspace*{0.2cm} (i,j=1,2).\nonumber\\
\label{eq:condition}
\eea

\subsection{Two-points functions in the asymptotic limit:}
\label{subsec:2puntos}
\hspace*{0.5cm}In order to present our results we write the functions 
$\Gamma^{V_{1}\, V_{2}}_{\mu\, \nu} (k)$ as,
\begin{equation}
\displaystyle \Gamma_{\mu \,\nu}^{V_{1} \, V_{2}}=
{\Gamma_{0}}_{\mu \,\nu}^{V_{1} \, V_{2}}+
\Delta {\Gamma}_{\mu \,\nu}^{V_{1} \, V_{2}}\,\,,
\end{equation}
where the tree level functions ${\Gamma_{0}}_{\mu \,\nu}^{V_{1} \, V_{2}}$ are given in 
eq.(\ref{eq:gam}) and the contributions from sfermions 
$\Delta {\Gamma}_{\mu \,\nu}^{V_{1} \, V_{2}}$ are defined by,
\begin{equation}
\Delta {\Gamma}_{\mu \,\nu}^{V_{1} \, V_{2}} (k) = 
\Sigma^{V_{1} \, V_{2}} (k)  g_{\mu\, \nu} + 
R^{V_{1} \, V_{2}} (k) k_{\mu} k_{\nu}\,\,.
\end{equation}
\hspace*{0.5cm}We omit to write the explicit formulae for the 
$\Sigma^{{\scriptscriptstyle XY}} (k)$ and $R^{{\scriptscriptstyle XY}} (k)$
 functions for brevity. The complete results can be found in~\cite{GEISHA}. 
 The results for the  $R^{{\scriptscriptstyle XY}} (k)$ functions
can, generically, be written as:
\begin{equation}
\label{eq:trampa}
R^{{\scriptscriptstyle XY}} (k) = -\left[\hspace*{0.1cm} k^{2}\hspace*{0.1cm}{\rm term}\hspace*{0.1cm}
 {\rm of}\hspace*{0.1cm}\Sigma^{{\scriptscriptstyle XY}} (k) \right] / k^{2}
\end{equation} 
\hspace*{0.5cm}As can be seen from our work in ref.\cite{GEISHA}, 
the asymptotic results in the large SUSY masses limit are of the generic form:
\bea
\Sigma^{V_{1} V_{2}} (k)&=&\Sigma^{V_{1} V_{2}}_{(0)}+
\Sigma^{V_{1} V_{2}}_{(1)} k^{2}+H\left[O\left(\frac{k^{2}}{\tilde{m}^{2}_{i}},
\frac{\tilde{m}_{i}^{2} - \tilde{m}_{j}^{2}}{\tilde{m}_{i}^{2} 
+ \tilde{m}_{j}^{2}}\right)\right] ,\nonumber\\
R^{V_{1} V_{2}} (k)&=&R^{V_{1} V_{2}}_{(0)}+J\left[O\left(\frac{k^{2}}{\tilde{m}^{2}_{i}},
\frac{\tilde{m}_{i}^{2} - \tilde{m}_{j}^{2}}{\tilde{m}_{i}^{2} 
+ \tilde{m}_{j}^{2}}\right)\right]\,\,,
\eea
where $\Sigma^{V_{1}\, V_{2}}_{(1)}$ and $R^{V_{1}\, V_{2}}_{(0)}$ contain the divergent 
$O(1/{\epsilon})$ contribution of dimensional regularization and are 
functions of the large SUSY masses but are $k$ independent. $\Sigma^{V_{1}\, V_{2}}_{(0)}$
is also a finite and k independent function, but not contains divergent contribution.
It goes to zero in our asymptotic behaviour. $H$ and $J$ are
finite functions which vanish in the large masses limit. 

In order to clarify the before comments, we present here the results for 
$\Sigma^{V_{1} V_{2}}_{(1)}$,
\bea
\label{eq:SumqAA}
\displaystyle {\Sigma_{\scriptscriptstyle (1)}^{A A}}_{\tilde{q}}(k)&=& 
- \frac{e^{2}}{16 \pi^{2}} \frac{N_{c}}{27} \sum_{\tilde{q}}
\left\{10 \Delta_{\epsilon} -4\log \frac{\tilde{m}_{t_{1}}^{2}}{\mu_{o}^{2}}\right.\nonumber\\ 
&& -4\log \frac{\tilde{m}_{t_{2}}^{2}}{\mu_{o}^{2}} \left. 
-\log \frac{\tilde{m}_{b_{1}}^{2}}{\mu_{o}^{2}}-\log
\frac{\tilde{m}_{b_{2}}^{2}}{\mu_{o}^{2}}
\right\},
\eea
\bea
\label{eq:SumqAZ}
\displaystyle &&{\Sigma_{\scriptscriptstyle (1)}^{A Z}}_{\tilde{q}}(k)= 
- \frac{e^{2}}{16 \pi^{2}} \frac{N_{c}}{9 s_{{\scriptscriptstyle W}} c_{{\scriptscriptstyle W}}} 
\sum_{\tilde{q}} \left\{ \left(\frac{3}{2}-\frac{10}{3}s_{{\scriptscriptstyle W}}^{2}\right)
\Delta_{\epsilon}\right.\nonumber\\
&&\hspace*{0.1cm}-\left(c_{t}^{2} - \frac{4}{3} s_{{\scriptscriptstyle W}}^{2}\right)
\log \frac{\tilde{m}_{t_{1}}^{2}}{\mu_{o}^{2}} -
\left(s_{t}^{2} - \frac{4}{3} s_{{\scriptscriptstyle W}}^{2}\right) 
\log \frac{\tilde{m}_{t_{2}}^{2}}{\mu_{o}^{2}}\nonumber\\
&& \left.
-\left(\frac{1}{2} c_{b}^{2} - \frac{1}{3} s_{{\scriptscriptstyle W}}^{2}\right) 
\log \frac{\tilde{m}_{b_{1}}^{2}}{\mu_{o}^{2}}-
\left(\frac{1}{2} s_{b}^{2} - \frac{1}{3} s_{{\scriptscriptstyle W}}^{2}\right) 
\log \frac{\tilde{m}_{b_{2}}^{2}}{\mu_{o}^{2}} \right\},\nonumber\\ 
\eea
\bea
\label{eq:SumqZZ}
\displaystyle &&{\Sigma_{\scriptscriptstyle (1)}^{Z Z}}_{\tilde{q}}(k)
= -\frac{e^{2}}{16\pi^{2}} \frac{N_{c}}
{3s_{{\scriptscriptstyle W}}^{2} c_{{\scriptscriptstyle W}}^{2}} \sum_{\tilde{q}} 
\left\{\frac{1}{18}(32s_{{\scriptscriptstyle W}}^{4}-
24s_{{\scriptscriptstyle W}}^{2}+9)\Delta_\epsilon \right.\nonumber\\
&&-\left(\frac{c_{t}^{2}}{2} - \frac{2 s_{{\scriptscriptstyle W}}^{2}}{3}\right)^{2} 
\log \frac{\tilde{m}_{t_{1}}^{2}}{\mu_{o}^{2}}  
-\left(\frac{s_{t}^{2}}{2} - \frac{2 s_{{\scriptscriptstyle W}}^{2}}{3}\right)^{2} 
\log \frac{\tilde{m}_{t_{2}}^{2}}{\mu_{o}^{2}}\nonumber\\
 && \displaystyle
-\left(-\frac{c_{b}^{2}}{2} + \frac{s_{{\scriptscriptstyle W}}^{2}}{3}\right)^{2}
\log \frac{\tilde{m}_{b_{1}}^{2}}{\mu_{o}^{2}}-
\left(-\frac{s_{b}^{2}}{2} + \frac{s_{{\scriptscriptstyle W}}^{2}}{3}\right)^{2}
\log \frac{\tilde{m}_{b_{2}}^{2}}{\mu_{o}^{2}}\nonumber \\
 && \displaystyle \left. -\frac{1}{2} s_{t}^{2} c_{t}^{2} 
\log \frac{\tilde{m}_{t_{1}}^{2} + \tilde{m}_{t_{2}}^{2}}
{2 \mu_{o}^{2}}- \frac{1}{2} s_{b}^{2} c_{b}^{2} 
\log \frac{\tilde{m}_{b_{1}}^{2} + \tilde{m}_{b_{2}}^{2}}{2 
\mu_{o}^{2}}\right\},
\eea
\bea
\label{eq:SumqWW}
\displaystyle {\Sigma_{\scriptscriptstyle (1)}^{W W}}_{\tilde{q}}(k)&=& 
- \frac{e^{2}}{16 \pi^{2}} \frac{N_{c}}{6s_{{\scriptscriptstyle W}}^{2}} 
\sum_{\tilde{q}} \left\{ 
\Delta_{\epsilon} -c_{t}^{2} c_{b}^{2} 
\log \frac{\tilde{m}_{t_{1}}^{2} + \tilde{m}_{b_{1}}^{2}}{2 \mu_{o}^{2}}
\right. \nonumber \\
\displaystyle &-& c_{t}^{2} s_{b}^{2} 
\log \frac{\tilde{m}_{t_{1}}^{2} + \tilde{m}_{b_{2}}^{2}}
{2 \mu_{o}^{2}} - s_{t}^{2} c_{b}^{2} 
\log \frac{\tilde{m}_{t_{2}}^{2} + \tilde{m}_{b_{1}}^{2}}
{2 \mu_{o}^{2}} \nonumber\\
&-&\left.  s_{t}^{2} s_{b}^{2} 
\log \frac{\tilde{m}_{t_{2}}^{2} + \tilde{m}_{b_{2}}^{2}}
{2 \mu_{o}^{2}}\right\}\,\,, 
\eea
where $\sw^{2}=sin^{2}{\theta}_{\scriptscriptstyle W}$ and $c_{f}=cos\theta_{f}, 
s_{f}=sin\theta_{f}$, with $\theta_{f}$ being the mixing angle in the $f$-sector.

Here and from now on,
\begin{equation}
\label{eq:lastint}
\hspace*{0.6cm}\displaystyle {\Delta}_\epsilon=\frac{2}{\epsilon }-{\gamma }_{\epsilon} 
+\log (4\pi) \hspace*{0.2cm}, \hspace*{0.2cm} \epsilon = 4-D\,,
\end{equation}
and $\mu_{o}$ is the usual mass scale of dimensional regularization.

As can be easily shown, it implies that all non-decoupling effects in the two-point
functions are contained in
$\Sigma^{V_{1}\, V_{2}}_{(1)}$ and $R^{V_{1}\, V_{2}}_{(0)}$ and, therefore, they can be 
absorbed into a redefinition of the SM relevant para\-meters, $m_{\scriptscriptstyle W},
m_{\scriptscriptstyle Z}$ and $e$ and the gauge bosons wave functions. In consequence, the
decoupling of squarks in the two point functions does indeed occur.

\subsection{Three-points functions in the asymptotic limit:}
\label{subsec:3puntos}
The result for the effective action under the mass conditions given in 
section~\ref{section:limit} can be written as,
\bea
\label{eq:efflim3}
&&{\Gamma_{eff}^{\tilde{f}} [V]}_{[3]} = \frac{1}{9{\pi}^2} \int 
 {\rm d\tilde{p}} \,\, {\rm d\tilde{q}} \,\, {\rm d\tilde{k}} \,\, \delta(p+k+q)\nonumber\\
 && \hspace*{0.7cm}\sum_{\tilde{f}} \left\{ \sum_{a,b,c} (\hat{O}_{p}^{1\,\mu})_{ab}
(\hat{O}_{\,q}^{1\,\nu})_{bc} (\hat{O}_{\,k}^{1\,\sigma})_{ca} \right.\,\nonumber\\
&& \hspace*{0.7cm} \left(\left.
{\Delta}_\epsilon-\log \frac{\mfa+\mfb+\mfc}{3\mu_{o}^{2}}\right) \,
{\L}_{\mu\,\nu\,\sigma}\right\} \,,\nonumber\\
\eea
where ${\L}_{\mu\,\nu\,\sigma}$ denotes the tree level operator defined by,
\begin{equation}
{\L}_{\mu\,\nu\,\sigma}\equiv\left[(p-q)_{\sigma}g_{\mu\, \nu}+
(k-p)_{\nu}g_{\mu\, \sigma}+(q-k)_{\mu}g_{\nu\, \sigma}\right]\,.
\end{equation}

Notice that this asymptotic result is proportional to the tree level operator 
${\L}_{\mu\,\nu\,\sigma}$ and therefore, we can at this point already conclude that the 
sfermions decouple in the three-point functions. We find interesting anyway to give 
explicitly also each contribution different from zero to the three-point Green's 
functions with specific external 
gauge bosons, $\Gamma_{\mu \,\nu \,\sigma}^{V_{1} V_{2} V_{3}}$. We present the results 
in the following form:
\begin{equation}
\displaystyle \Gamma_{\mu \,\nu \,\sigma}^{V_{1} \, V_{2} \, V_{3}}=
{\Gamma_{0}}_{\mu \,\nu \,\sigma}^{V_{1} \, V_{2} \, V_{3}}+
\Delta {\Gamma}_{\mu \,\nu \,\sigma}^{V_{1} \, V_{2} \, V_{3}}\,,
\end{equation}
where the contributions at tree level are:
\begin{equation}
\displaystyle {\Gamma_{0}}_{\mu \,\nu \,\sigma}^{A W^+ W^-}=-e\,{\L}_{\mu \nu \sigma}\,\,,
\,\,\, {\Gamma_{0}}_{\mu \,\nu \,\sigma}^{Z W^+ W^-}=-g\cw\,{\L}_{\mu \nu \sigma}\,.
\end{equation}
\hspace*{0.5cm}In order to get the sfermions contributions, one must substitute all the "operators" that 
appear in eq.(\ref{eq:efflim3}), perform the corresponding sums and after rather lengthy
calculation, the following results, written in a compact form, are obtained:
\bea
\label{eq:AWW}
\displaystyle {\Delta {\Gamma}_{\mu \,\nu \,\sigma\hspace*{0.4cm}\tilde{q}}^{A W^+ W^-}}&=&
\frac{eg^{2}}{16 \pi^{2}} \frac{N_{c}}{9}\, {\L}_{\mu\,\nu\,\sigma} \sum_{\tilde{q}}
\frac{1}{2} \left\{\left(\Delta_{\epsilon} +
\log \mu_{o}^{2}\right)\right.\nonumber\\
&+&\left.f_{1}(\tilde{m}_{t_{1}}^{2}, \tilde{m}_{t_{2}}^{2}, \tilde{m}_{b_{1}}^{2},
\tilde{m}_{b_{2}}^{2})\right\} \nonumber\\
&+& \displaystyle {F_{1}}_{\mu \,\nu \,\sigma}\left[
O\left(\frac{p{2}}{\tilde{m}^{2}},\,\frac{\tilde{m}_{i}^{2} - 
\tilde{m}_{j}^{2}}{\tilde{m}_{i}^{2} 
+ \tilde{m}_{j}^{2}}\right)\right]\,\,,
\eea
\bea
\label{eq:ZWW}
\displaystyle {\Delta {\Gamma}_{\mu \,\nu \,\sigma\hspace*{0.4cm}\tilde{q}}^{Z W^+ W^-}}&=&
 -\frac{g^{3}}{16 \pi^{2}} \frac{N_{c}}{6\cw}\,{\L}_{\mu\,\nu\,\sigma} 
\sum_{\tilde{q}}\frac{1}{3}s_{{\scriptscriptstyle W}}^{2} 
\left\{\left(\Delta_{\epsilon} +\log \mu_{o}^{2}\right)\right.\nonumber\\
&+&\left.f_{2}(\tilde{m}_{t_{1}}^{2}, \tilde{m}_{t_{2}}^{2}, \tilde{m}_{b_{1}}^{2},
\tilde{m}_{b_{2}}^{2})\right\} \nonumber\\
&+& \displaystyle {F_{2}}_{\mu \,\nu \,\sigma}\left[
O\left(\frac{p{2}}{\tilde{m}^{2}},\,\frac{\tilde{m}_{i}^{2} - 
\tilde{m}_{j}^{2}}{\tilde{m}_{i}^{2} 
+ \tilde{m}_{j}^{2}}\right)\right]\,,
\eea
where the functions ${F_{i}}_{\mu \,\nu \,\sigma} \,(i=1,2)$ are finite and we have proved
explicitly that they go to zero in the limit of $\tilde{m}\rightarrow\infty$ with 
$|\tilde{m}_{i}^{2} - \tilde{m}_{j}^{2}| \ll
|\tilde{m}_{i}^{2} + \tilde{m}_{j}^{2}|$.

In the above two expressions, the functions $f_{i}(\tilde{m}_{t_{1}}^{2}, \tilde{m}_{t_{2}}^{2},
\tilde{m}_{b_{1}}^{2}, \tilde{m}_{b_{2}}^{2}) \, (i=1,2)$ are finite and different from zero 
in the large masses limit. Therefore, they contain all the potentially 
non-decoupling effects of the three-point functions. Their explicit expressions can be 
found in \cite{DMS}. In principle, the dependence on the various sfermions masses of each
of these functions are diffe\-rent from each other. However, in order to implement the 
large supersymmetric masses limit 
one must choose a proper combination of masses such that there is just one large mass 
parameter. We choose here, suitably, a sum of three masses as the large parameter.
The remaining mass parameters can be expressed in terms of the differences
of masses which in our approximation are small as compared to the sum. In terms of 
these mass combination, we get:
\bea
f_{1}(\tilde{m}_{t_{1}}^{2}, \tilde{m}_{t_{2}}^{2}, \tilde{m}_{b_{1}}^{2},
\tilde{m}_{b_{2}}^{2}) &=& -\log \hat{M}^{2}_{1}+
O\left(\frac{\tilde{m}_{i}^{2} - \tilde{m}_{j}^{2}}{\hat{M}^{2}}\right)\,,\nonumber\\
f_{2}(\tilde{m}_{t_{1}}^{2}, \tilde{m}_{t_{2}}^{2}, 
\tilde{m}_{b_{1}}^{2}, \tilde{m}_{b_{2}}^{2}) &=& -\log \hat{M}^{2}_{2}+
O\left(\frac{\tilde{m}_{i}^{2} - \tilde{m}_{j}^{2}}{\hat{M}^{2}}\right)\,,\nonumber\\
\eea
being $\hat{M}^{2}_{1}=\frac{1}{3}\left(\tilde{m}_{t_{2}}^{2}+
2\tilde{m}_{b_{1}}^{2}\right)$ and $\hat{M}^{2}_{2}=\frac{1}{3}
\left(2\tilde{m}_{t_{2}}^{2}+\tilde{m}_{b_{1}}^{2}\right).$

As we have mentioned above, the corrections $\Delta {\Gamma}$ are proportional to the 
tree level, ${\L}_{\mu\,\nu\,\sigma}$, and therefore the potentially non-decoupling 
effects in the three-point functions can be absorbed into redefinitions of the coupling 
constants and wave functions. 

\subsection{Four-points functions in the asymptotic limit:}
Analogously to the previous section, we write the results for the four-point functions as,
\begin{equation}
\displaystyle \Gamma_{\mu \,\nu \,\sigma\,\lambda}^{V_{1} \, V_{2} \, V_{3}\, V_{4}}=
{\Gamma_{0}}_{\mu \,\nu \,\sigma\,\lambda}^{V_{1} \, V_{2} \, V_{3}\, V_{4}}+
\Delta {\Gamma}_{\mu \,\nu \,\sigma\,\lambda}^{V_{1} \, V_{2} \, V_{3}\, V_{4}}\,,
\end{equation}
where the different contributions to the effective action at tree level
are defined by,
\begin{eqnarray}
\displaystyle \Gamma_{0\,\,\mu \,\nu \,\sigma\,\lambda}^{AAW^+ W^-}=-e^{2}
{\ss}_{\mu \nu \sigma \lambda}&,&
\Gamma_{0\,\,\mu \,\nu \,\sigma\,\lambda}^{AZW^+W^-}=
-g^{2}s_{w}c_{w}{\ss}_{\mu \nu \sigma \lambda}, \nonumber\\
\nonumber\\
\displaystyle \Gamma_{0\,\,\mu \,\nu \,\sigma\,\lambda}^{ZZW^+ W^-}=
-g^{2}\cw{\ss}_{\mu \nu \sigma \lambda}&,&
\Gamma_{0\,\,\mu \,\nu \,\sigma\,\lambda}^{W^+ W^-W^+W^-}=
g^{2}{\ss}_{\mu \nu \sigma \lambda}\,\,,\nonumber\\
\end{eqnarray}
with,
\begin{equation}
{\ss}_{\mu\,\nu\,\sigma\,\lambda}\equiv\left[ 2g_{\mu \,\nu}g_{\sigma \,\lambda}-
g_{\mu \,\sigma}g_{\nu \,\lambda}-g_{\mu \,\lambda}g_{\nu \,\sigma}\right]\,.
\end{equation}

Similar expressions to the eqs.(\ref{eq:AWW}) and (\ref{eq:ZWW}) are obtained for the squarks 
contributions to the four-point functions,
\begin{eqnarray}
\label{eq:AAWW}
\displaystyle {\Delta {\Gamma}_{\mu \,\nu \,\sigma\,\lambda\hspace*{0.4cm}\tilde{q}}
^{A A W^+ W^-}}&=&
-\frac{N_{c}}{6} \frac{e^{2}g^{2}}{16 \pi^{2}} \,{\ss}_{\mu \nu \sigma \lambda} 
\,\sum_{\tilde{q}}\left\{\left(\Delta_{\epsilon} +\log \mu_{o}^{2}\right)\right.\nonumber\\
&+& \left.
g_{1}(\tilde{m}_{t_{1}}^{2}, \tilde{m}_{t_{2}}^{2}, \tilde{m}_{b_{1}}^{2},
\tilde{m}_{b_{2}}^{2})\right\} \nonumber\\
&+& \displaystyle {G_{1}}_{\mu \,\nu \,\sigma\,\lambda}\left[
O\left(\frac{p{2}}{\tilde{m}^{2}},\,\frac{\tilde{m}_{i}^{2} - 
\tilde{m}_{j}^{2}}{\tilde{m}_{i}^{2}+\tilde{m}_{j}^{2}}\right)\right]\,\,,\nonumber\\
\end{eqnarray}
\begin{eqnarray}
\label{eq:AZWW}
\displaystyle {\Delta {\Gamma}_{\mu \,\nu \,\sigma\,\lambda\hspace*{0.4cm}\tilde{q}}
^{A Z W^+ W^-}}&=&-\frac{N_{c}}{6} 
\frac{eg^{3}}{16 \pi^{2}}c_{\scriptscriptstyle W} \,{\ss}_{\mu \nu \sigma \lambda} 
\sum_{\tilde{q}}\left\{\left(\Delta_{\epsilon} +\log \mu_{o}^{2}\right)
\right.\nonumber\\
&+&\left.g_{2}(\tilde{m}_{t_{1}}^{2}, \tilde{m}_{t_{2}}^{2}, \tilde{m}_{b_{1}}^{2},
\tilde{m}_{b_{2}}^{2})\right\} \nonumber\\
&+& \displaystyle {G_{2}}_{\mu \,\nu \,\sigma\,\lambda}\left[
O\left(\frac{p{2}}{\tilde{m}^{2}},\,\frac{\tilde{m}_{i}^{2} - 
\tilde{m}_{j}^{2}}{\tilde{m}_{i}^{2}+\tilde{m}_{j}^{2}}\right)\right]\,\,,\nonumber\\
\end{eqnarray}
\begin{eqnarray}
\label{eq:ZZWW}
\displaystyle {\Delta {\Gamma}_{\mu \,\nu \,\sigma\,\lambda\hspace*{0.4cm}\tilde{q}}
^{Z Z W^+ W^-}}&=&
-\frac{N_{c}}{6} \frac{g^{4}}{16 \pi^{2}}c_{\scriptscriptstyle W}^{2} 
\,{\ss}_{\mu \nu \sigma \lambda} 
\sum_{\tilde{q}}\left\{\left(\Delta_{\epsilon} +\log \mu_{o}^{2}\right)
\right.\nonumber\\
&+&\left.g_{3}(\tilde{m}_{t_{1}}^{2}, \tilde{m}_{t_{2}}^{2}, \tilde{m}_{b_{1}}^{2},
\tilde{m}_{b_{2}}^{2})\right\} \nonumber\\
&+& \displaystyle {G_{3}}_{\mu \,\nu \,\sigma\,\lambda}\left[
O\left(\frac{p{2}}{\tilde{m}^{2}},\,\frac{\tilde{m}_{i}^{2} - 
\tilde{m}_{j}^{2}}{\tilde{m}_{i}^{2}+\tilde{m}_{j}^{2}}\right)\right]\,\,,\nonumber\\
\end{eqnarray}
\begin{eqnarray}
\label{eq:WWWW}
\displaystyle {\Delta {\Gamma}_{\mu \,\nu \,\sigma\,\lambda\hspace*{0.9cm}\tilde{q}}
^{W^+ W^- W^+ W^-}}&=&
-\frac{N_{c}}{3}\frac{g^{4}}{16 \pi^{2}}\,{\ss}_{\mu \nu \sigma \lambda} 
\sum_{\tilde{q}}\left\{\left(\Delta_{\epsilon} +\log \mu_{o}^{2}\right)
\right.\nonumber\\
&+& \left.g_{4}(\tilde{m}_{t_{1}}^{2}, \tilde{m}_{t_{2}}^{2}, \tilde{m}_{b_{1}}^{2},
\tilde{m}_{b_{2}}^{2})\right\} \nonumber\\
&+& \displaystyle {G_{4}}_{\mu \,\nu \,\sigma\,\lambda}\left[
O\left(\frac{p{2}}{\tilde{m}^{2}},\,\frac{\tilde{m}_{i}^{2} - 
\tilde{m}_{j}^{2}}{\tilde{m}_{i}^{2}+\tilde{m}_{j}^{2}}\right)\right]\,,\nonumber\\
\end{eqnarray}
\hspace*{0.5cm}It is important to point out that the functions ${G_{k}}_{\mu \,\nu \,\sigma\,\lambda}$
and $g_{k}(\tilde{m}_{t_{1}}^{2}, \tilde{m}_{t_{2}}^{2}, \tilde{m}_{b_{1}}^{2},
\tilde{m}_{b_{2}}^{2}) \,(k=1,\ldots4)$ are both finite, but the first ones go to zero in our asymptotic
limit whereas the second ones are different from zero in this limit, and therefore these
latter contain all the potentially non-decoupling effects of the four-point functions. 
For brevity, we will not present 
here the explicit formulae of $g_{k}$ functions. However, we would like to have noticed that if one 
takes the sum of all the masses as the large parameter in the expansion of the 
logarithm's coefficients one get,
\begin{equation}
g_{k}(\tilde{m}_{t_{1}}^{2}, \tilde{m}_{t_{2}}^{2}, \tilde{m}_{b_{1}}^{2},
\tilde{m}_{b_{2}}^{2}) \,=-\log \hat{M}^{2}+
O\left(\frac{\tilde{m}_{i}^{2} - \tilde{m}_{j}^{2}}{\hat{M}^{2}}\right)\,,
\end{equation}
where $\hat{M}^{2}=\frac{1}{4}(\tilde{m}_{t_{1}}^{2}+\tilde{m}_{t_{2}}^{2}+
\tilde{m}_{b_{1}}^{2}+\tilde{m}_{b_{2}}^{2})$.

From eqs.(\ref{eq:AAWW}) through (\ref{eq:WWWW}), it can be seen that all the corrections 
$\Delta {\Gamma}$ are proportional to the 
tree level, ${\ss}_{\mu \nu \sigma \lambda}$, and therefore the potentially 
non-decoupling effects in the four-point functions can be also absorbed into redefinitions
of the coupling constants and wave functions. 
 
Similar results are obtained for the sleptons sector for Green's functions doing the
corresponding replacements: $\tilde{q} \rightarrow \tilde{l},\,
N_{c} \rightarrow 1,\, \tilde{m}_{t_{1}} \rightarrow  \tilde{m}_{\nu},\,
\tilde{m}_{b_{1}} \rightarrow  \tilde{m}_{\tau_{1}},\, 
\tilde{m}_{b_{2}} \rightarrow  \tilde{m}_{\tau_{2}},\, 
c_{t} \rightarrow 1, s_{t} \rightarrow 0, c_{b} \rightarrow c_{\tau}$ and
$s_{b} \rightarrow s_{\tau}\,.$ 

In summary, from our results it is
clear that there is indeed decoupling in the two, three and four-point electroweak gauge boson 
functions:\\
All SUSY effects can be absorbed into redefinitions of $m_{\scriptscriptstyle Z},
m_{\scriptscriptstyle W}, e$ and the wave functions of the gauge bosons $W^{\pm}, Z, A$, or
else they are suppressed by inverse powers of the heavy SUSY particles masses.

\section{Conclusions}
\label{sec:con}
\hspace*{0.5cm} The computation of the effective action for the standard particles which results
by integrating out all the heavy supersymmetric particles will provide the answer
to the question whether the decoupling of heavy supersy\-mmetric particles in the MSSM occurs
leading to the SM as the remaining low energy effective theory. In this work we have 
shown that all the contributions from the heavy sfermions to the two, three and four-point 
functions of the electroweak gauge bosons can be absorbed into redefinitions of the 
SM parameters or they are suppressed by inverse powers of the heavy sparticles masses. 
Therefore we have proved analytically that the decoupling of heavy sfermions at one loop
level does occur.

We have considered the asymptotic limit where
the sfermion masses are all large as compared to the $W^{\pm}$ and $Z$
masses and the external momenta and we have always worked under the assumption 
that the differences of their squared masses are much smaller than their sums. Notice that we 
have not assumed exact universality of the masses.

Our results for these Green functions in the large SUSY masses limit have 
been presented analytically and given in terms of the sparticle masses. They do not depend
on the particular choice for the soft-breaking terms and therefore they are general
results. In our opinion, it is more convenient for 
the analysis of the phenomenon  of decoupling to use the physical sparticle masses
as, being the proper parameters, rather than some other possible mass parameters 
of the MSSM as, for instance, the $\mu$-parameter or the soft-SUSY breaking 
parameters.
     
Finally, we have explored to what extent the hypothe\-sis of generation of SUSY masses by 
soft-SUSY breaking terms is relevant for decoupling and we have found instead that the 
requirement of $SU(3)_{\rm c} \times \gs$ gauge invariance of the explicit mass terms by itself is 
sufficient to get it.

A complete proof of decoupling of supersymmetric particles will include the integration 
of all the heavy supersymmetric spectrum. Here we only discuss the squarks and sleptons sectors.
The analysis of charginos and neutralinos have been done in \cite{GEISHA,DMS}. The 
integration of the Higgs sector will be considered in a forthcoming work.\cite{NEW}
The complete proof would lead eventually to the conclusion that the SM is indeed the low energy effective 
of the MSSM in the large SUSY masses limit.

\section*{Acknowledgements.}
M.J.H wish to acknowledge the organizers of the International Conference on High Energy Physics 98
 for their kind hospitality during the school. This 
paper is based in the previous work finished in the last months \cite{GEISHA,DMS}, which has 
been partially supported by the Spanish Mi\-nisterio de Educaci{\'o}n y Ciencia 
under projects CICYT AEN96-1664 and AEN93-0776, and the fellowship AP95 00503301.

\section*{Appendix A.}
\vspace{0.4cm}
\setcounter{equation}{0}
\renewcommand{\theequation}{A.\arabic{equation}}

In this appendix we give the definition of the one-loop integrals 
$I_{0}\,, I^{a\,b}_{f_{\mu\, \nu}}\,,T^{a\,b}_{\mu}\,,
T^{a\,b\, c}_{\mu \,\nu \,\sigma}\,,J^{a\,b}\,,
J^{a\,b\, c}_{\mu \,\nu}$ and $J^{a\,b\,c\,d}_{\mu \,\nu\, \sigma\,\lambda}$
that have been used in the computation of the two, three and
four-point functions and their results in the large masses limit.
We start by giving the definition of the integrals in dimensional regularization.\\

$\bullet$ One loop integrals.
\bea
\label{eq:int0}
I_{0}&=&  \int d\widehat{q} \frac{1}{\left[q^2 - \mfa \right]} \,,\\
I^{a\,b}_{f_{\mu\, \nu}}&=&  \int d\widehat{q} 
\frac{(2q+k)_{\mu} (2q+k)_{\nu}}
{\left[(k+q)^2 - \tilde{m}^2_{f_a}\right] \left[q^2 - \tilde{m}^2_{f_b}\right]}\,,\\
\label{eq:int23}
T^{a\,b}_{\mu} &=&\int d\widehat{p} \,\frac{(2p+q)_{\mu}}
{\left[p^2 - \tilde{m}^2_{f_a}\right]\left[(p+q)^2 - \tilde{m}^2_{f_b}\right]}\,\,,
\eea
\bea
&&T^{a\,b\, c}_{\mu \,\nu \,\sigma} =8I^{a\,b\, c}_{\mu \,\nu \,\sigma}+
4 [k_{\sigma}I^{a\,b\, c}_{\mu \,\nu}+ (k-p)_{\nu}I^{a\,b\, c}_{\mu \,\sigma}+
p_{\mu}I^{a\,b\, c}_{\nu \,\sigma}] \nonumber\\
&& +2 [k_{\sigma}(k-p)_{\nu}I^{a\,b\, c}_{\mu}+k_{\sigma}p_{\mu}I^{a\,b\, c}_{\nu}-
(k-p)_{\nu}p_{\mu}I^{a\,b\, c}_{\sigma}]\nonumber\\
&&+p_{\mu}k_{\sigma}(k-p)_{\nu}I^{a\,b\, c}\,\,,
\eea
where,
\bea
I^{a\,b\, c}=\int d\widehat{p} \,\,\frac{1}{\it D}\,\,\,&,&\,
I^{a\,b\, c}_{\mu}=\int d\widehat{p} \,\,
\frac{p_{\mu}}{\it D}\,\,,\nonumber\\
I^{a\,b\, c}_{\mu\,\nu}=\int d\widehat{p} \,\,
\frac{p_{\mu}\,p_{\nu}}{\it D}\,\,\,&,&\,
I^{a\,b\, c}_{\mu\,\nu\,\sigma}=\int d\widehat{p} \,\,
\frac{p_{\mu}\,p_{\nu}p_{\sigma}}{\it D}\,,\nonumber\\
\eea
defining {\it D} as,\\
\begin{center}
${\it D}=\left[p^2 - \tilde{m}^2_{f_a}\right]\left[(p+q)^2 - \tilde{m}^2_{f_b}\right]
\left[(p+q+k)^2 - \tilde{m}^2_{f_c}\right]\,.$\\
\end{center}
\vspace{0.3cm}
And,
\bea
J^{a\,b}&=&\int d\widehat{p} \,
\frac{1}{\left[p^2 - \tilde{m}^2_{f_a}\right]\left[(p+q+k)^2 - \tilde{m}^2_{f_b}
\right]}\,,\\
J^{a\,b\, c}_{\mu\,\nu}&=&\int d\widehat{p}\,\,\,
\frac{{\it P}_{\mu \nu}}{\it E}\,\,,\\
J^{a\,b\,c\,d}_{\mu\,\nu\,\sigma \,\lambda}
&=& \int d\widehat{p} \,\, 
\frac{{\it P}_{\mu \nu}(2p+2q+2k+r)_{\sigma}(2p+q+k+r)_{\lambda}}
{{\it E}\left[(p+q+k+r)^2 - \tilde{m}^2_{f_d}\right]}\,\,,\nonumber\\
\eea
where,\\
\begin{center}
${\it E}=\left[p^2 - \tilde{m}^2_{f_a}\right]\left[(p+q)^2 - \tilde{m}^2_{f_b}\right]
\left[(p+q+k)^2 - \tilde{m}^2_{f_c}\right]$\\ 
\end{center}
\vspace{0.2cm}
\begin{center}
$\,{\it P}_{\mu \nu}=(2p+q)_{\mu} \,(2p+2q+k)_{\nu}$\\
\end{center}

$\bullet$ Results in the asymptotic limit.\\

The results in the large masses limit have been obtained using the m-Theorem \cite{GMR}.
\bea
\label{eq:io}
\displaystyle I_{0}&=& \frac{i}{16 \pi^{2}} 
\left( {\Delta}_\epsilon+1-\log \frac{\mfa}{\mu_{o}^{2}} \right) {\mfa}\nonumber \\
\displaystyle I^{ab}_{{\scriptscriptstyle f}} 
&=& \frac{i}{16 \pi^{2}} \left\{(\mfa+\mfb) \left({\Delta}_\epsilon+1
-\log \frac{\mfa+\mfb}{2\mu_{o}^{2}}\right)g_{\mu\, \nu}\right.\nonumber \\
&& -\frac{1}{3} k^{2}  \left({\Delta}_\epsilon-\log 
\frac{\mfa+\mfb}{2\mu_{o}^{2}}\right)g_{\mu\, \nu} \nonumber \\
&&+ \left.\frac{1}{3} k_{\mu}k_{\nu}\left({\Delta}_\epsilon-
\log \frac{\mfa+\mfb}{2\mu_{o}^{2}}\right)\right\}\nonumber \\
\nonumber \\
\displaystyle T^{a\,b}_{\mu}&=&0\,\,,\,\,
\displaystyle I^{a\,b\, c}=\,I^{a\,b\, c}_{\mu}\, = \, 0 \,,\nonumber \\
\nonumber \\
\displaystyle I^{a\,b\, c}_{\mu\,\nu} &=& \frac{i}{16 \pi^{2}} 
\frac{1}{4} \left({\Delta}_\epsilon-\log 
\frac{\mfa+\mfb+\mfc}{3\mu_{o}^{2}}\right)g_{\mu\, \nu}\,,\nonumber \\
\nonumber \\
\displaystyle I^{a\,b\, c}_{\mu\,\nu\,\sigma}&=& -\frac{i}{16 \pi^{2}}
\frac{1}{12}{(2q+k)}_{\rho}
\left({\Delta}_\epsilon-\log \frac{\mfa+\mfb+\mfc}{3\mu_{o}^{2}}\right)*\nonumber \\
&&\hspace*{0.3cm}\left[g_{\mu\, \sigma}g_{\nu\, \rho}+g_{\mu\, \rho}g_{\nu\, \sigma}
\right]\,,\nonumber \\
\nonumber \\
\displaystyle T^{a\,b\, c}_{\mu \,\nu \,\sigma} &=& \frac{i}{16 \pi^{2}}
\frac{1}{3}\left(
{\Delta}_\epsilon-\log \frac{\mfa+\mfb+\mfc}{3\mu_{o}^{2}}\right)*\nonumber \\
& & \displaystyle \left[(p-q)_{\sigma}g_{\mu\, \nu}+
(k-p)_{\nu}g_{\mu\, \sigma}+(q-k)_{\mu}g_{\nu\, \sigma}\right]\nonumber \\
\nonumber \\
\displaystyle J^{a\,b\, c}&=&
\frac{i}{16 \pi^{2}}\left({\Delta}_\epsilon-\log \frac{\mfa+\mfb}{2\mu_{o}^{2}}\right)\,,\nonumber \\
\nonumber \\
\displaystyle J^{a\,b\, c}_{\mu\,\nu}&=&\frac{i}{16 \pi^{2}} \left({\Delta}_\epsilon-
\log \frac{\mfa+\mfb+\mfc}{3\mu_{o}^{2}}\right)g_{\mu\, \nu}\,,
\end{eqnarray}
and finally,
\begin{eqnarray}
\label{eq:last4}
\displaystyle J^{a\,b\,c\,d}_{\mu\,\nu\,\sigma \,\lambda}&=&
\frac{i}{16 \pi^{2}}\frac{2}{3}\left(
{\Delta}_\epsilon-\log \frac{\mfa+\mfb+\mfc+\mfd}{4\mu_{o}^{2}}\right)*\nonumber\\
&&\left[g_{\mu\, \nu}g_{\sigma\, \lambda}+g_{\mu\, \sigma}g_{\nu\, \lambda}
+g_{\mu\, \lambda}g_{\nu\, \sigma}\right]\,.
\end{eqnarray}
The corrections to these formulae are suppressed by inverse powers of the sums of the 
sfermion masses and vanish in the large masses limit.

\end{document}